# EXCITATION OF QUASI-MONOCHROMATIC WAVES BY A HIGH-VOLTAGE PULSE IN A FERRITE COAXIAL LINE WITH THE PERIODIC STRUCTURE

*A.B. Batrakov, S.Yu. Karelin, O.M. Lebedenko, V.S. Mukhin, I.N. Onishchenko, O.L. Rak, V.G. Sinitsin, M.V. Volovenko*

*National Science Center «Kharkiv Institute of Physics and Technology», Ukraine*

*E-mail: sergeykarelin1976@gmail.com*

Experimental data and results of numerical simulations are presented, concerning excitation of narrowband gigahertz-range wave trains in coaxial guiding structures that are partially filled with ferromagnetic material and may involve periodically arranged metal inserts. The experiments performed confirm the possibility of exciting weakly damped electromagnetic waves by feeding high voltage, unilateral electromagnetic pulses of short duration into the line. The coax line was of outer diameter 50.5 mm, filled with an isotropic dielectric (relative dielectric constant $\varepsilon = 2.25$) and a set of ferrite rings with $\varepsilon=16$ and saturated-state $\mu$ about 4 to 5. With a peak voltage of the primary pulse close to 160 kV and a magnetizing field of 17.5 kA/m, the parameters of the waves excited reached magnitudes as**:** frequency 1.89 GHz to 2.1 GHz; bandwidth 16%; VHF power at the output about 20 MW.

PACS: 41.20 Gz; 41.20 Jb

## 1. INTRODUCTION

As of recent, the possibility of direct conversion of the energy of a high-voltage, unipolar pulse into high-frequency electromagnetic waves has been attracting a lot of interest [1-4]. The effect is expected to occur in the course of transformation of the shock wave generated by the primary pulse within the transmission line it propagates through (the line with nonlinear properties, NLPL). Specifically that should be a coaxial line partially filled with saturated ferrite. Meanwhile, the amplitude of the electromagnetic wave formed in this way decreases rather quickly and the period increases. As a result, the spectrum of the wave train may becomes very broad. In paper [5], it was suggested to provide for the formation of an electromagnetic wave with an appropriate dispersion law by placing the NLPL and some periodic structure (PS) on the line's central electrode, while in paper [6] on the inner wall of the outer electrode of cylindrical cross-section. Such arrangements should permit excitation of electromagnetic waves with a slow rate of decay and, accordingly, a narrow spectrum.

In paper [7], an optimization of the NLPL's design with a PS was carried out and some preliminary experimental results obtained that seemed to confirm, in principle, effectiveness of the method proposed. However, the oscillations obtained remained of low amplitude (only 10 to 15 kV in the case of pulse voltages about 200 kV). The purpose of this work is to suggest ways for increasing the amplitude and duration of the wave trains excited.

## 2. EXPERIMENT

In order to increase both the amplitude and duration of the oscillations (e.g., as compared with the data of paper [7]), the leading edge of the pulse was compressed and the length of the structure (NLPL with the PS) increased. A dual forming line (DFL) was suggested as the primary pulse source, and a pulse sharpening circuit was added, aimed at enriching the spectral content of the waveform with higher frequencies (in particular, through 'pre-sharpening' the pulse's front edge. As a result, an input pulse with a front of ~3 ns and a half-life duration of ~15 ns was obtained, which should make it possible to increase the field amplitude. To obtain longer lasting oscillations, the length of the NLPL was increased from 100 mm to 264 mm.

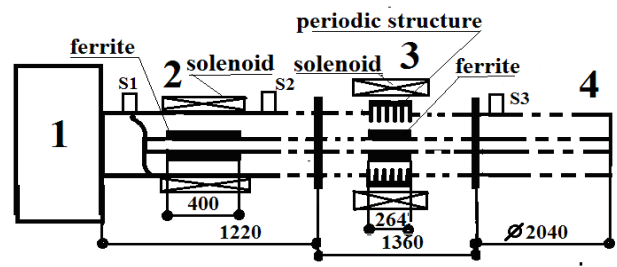

*Fig. 1. Setup scheme*

As a result, the experimental setup consisted of the following elements (Fig. 1):

1 – high-voltage (up to 160 kV) source of primary pulses – DFL;

2 – pulse front sharpener with a ferrite insert;

S1, S2, and S3 – voltage sensors on the outer electrode, which use capacitive voltage dividers;

3 – section with a metallic periodic structure and a ferrite insert;

4 – coaxial line with a load at the output;

As a result of using a ferrite containing pulse sharpener (2), the pulse front width could be reduced, compared with the figure of paper [7], by a factor of two or more, from 1 ns to 0.5 ns.

The 'compressed' pulse was fed from the PS (3) to the NLPL. Shown in Fig. 2a are the pulsed forms as they looked before and after passage through the PS-containing NLPL. The amplitude of the waveform excited can be seen to be A=32 kV, which, with the known line impedance Z = 36 Ohm, corresponds to the VHF power W ≈ 20 MW. The wave's frequency can be estimated from the spectrum of the pulse at the output (Fig. 2b), specifically $f_G$= 1.89 GHz, with a bandwidth $\Delta f/f_G$= 16%.

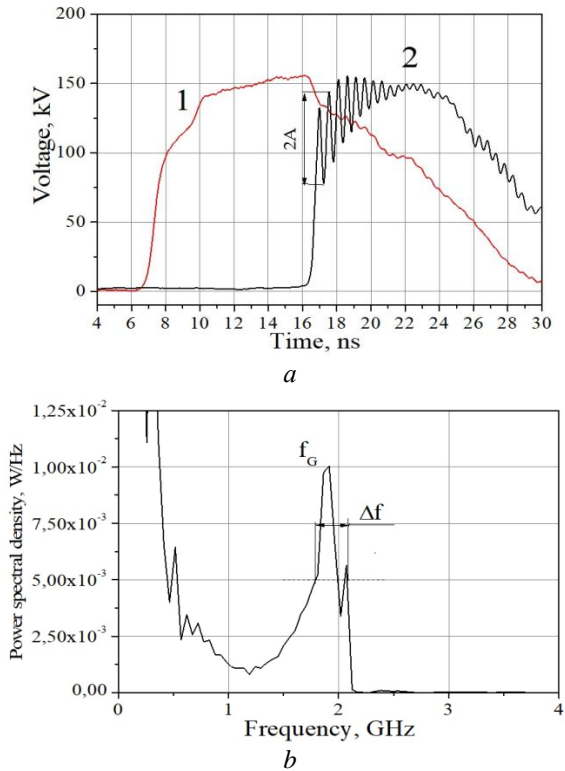

*Fig. 2. Pulsed waveform and its spectrum as observed in experiments*

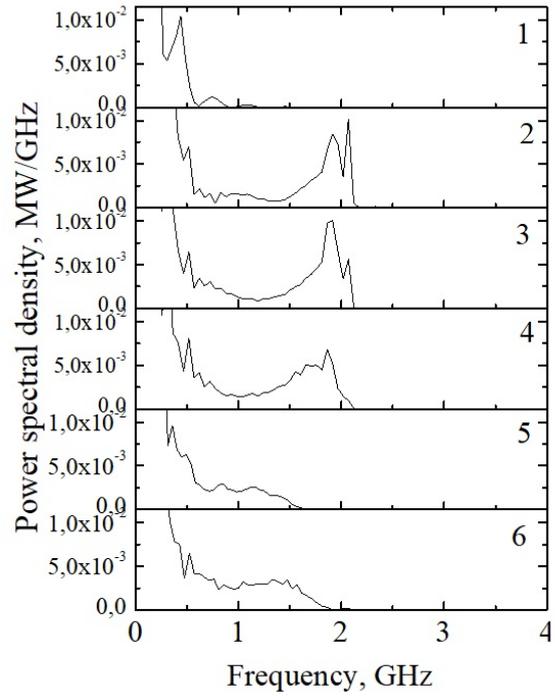

*Fig.3. The pulsed voltage waveform (panel a) for a variety of incoming pulse voltages:*

*1 – 160 kV, 2 – 134 kV, 3 – 107 kV, 4 – 80 kV, 5 – 67 kV; and its spectrum (panel b) as read by sensor S3 for a variety of incoming voltages: 1 -160 kV*

As follows from experimental data (see Fig. 3), the maximum amplitude and duration of the wave are reached when the value of the magnetizing field $H_0$ lies in the range of 17–35 kA/m. At lower magnitudes than $H_0$=10 kA/m (that is, when the ferrite is not in the state of full magnetic saturation), the oscillations disappeared abruptly. With the magnetizing field increasing to magnitudes above 35 kA/m the oscillations gradually grew shorter, with their amplitude decreasing.

It should also be noted that the largest oscillations were achieved at the highest input pulse voltage (Fig. 4) − 160 kV, although the numerical calculation in [8] predicted the presence of a certain optimum in the amplitude and duration of oscillations depending on the pulse voltage. It can be assumed that with an increase in the voltage of the input pulse above 160 kV, the amplitude of oscillations will increase even more. However, further increasing the voltage is associated with significant technical difficulties due to the increased risk of electrical breakdown.

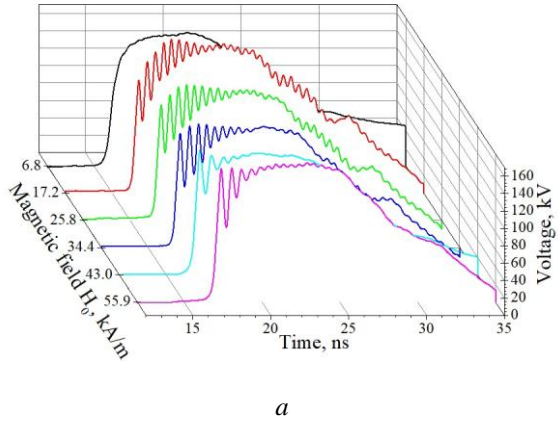

*a*

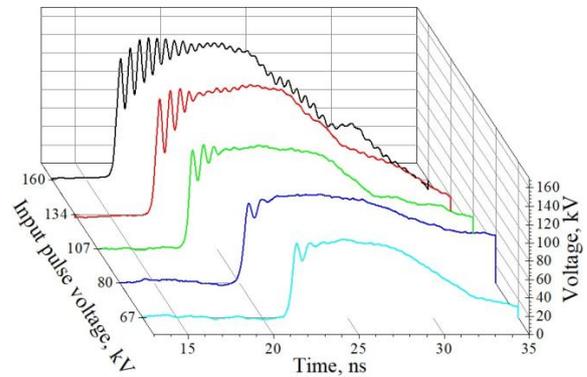

*a*

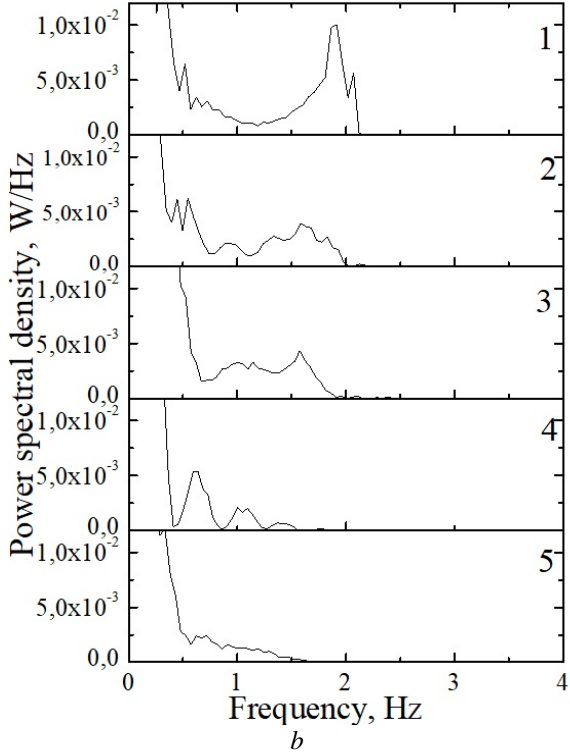

*Fig. 4. Voltage pulse forms at sensor S3 (a) and respective spectra (b) for different input voltages:*
*1 – 160 kV, 2 – 134 kV, 3 – 107 kV, 4 – 80 kV, 5 – 67 kV*

## 3. NUMERICAL CALCULATIONS AND DISCUSSION

For comparison with the experimental results, a numerical calculation was performed using the FDTD method according to [8]. The results of this calculation (Fig. 5) agree well with the experiment in the frequency of oscillations ~1.9 GHz. Also, the calculation adequately describes the gradual disappearance of oscillations with the growth of the saturating magnetic field $H_0$, which occurs due to the gradual exit of ferrite from the region of nonlinearity.

It should also be noted that the largest oscillations were achieved at the highest input pulse voltage (Fig. 4) – 160 kV, although the numerical calculation in [8] predicted the presence of a certain optimum in the amplitude and duration of oscillations depending on the pulse voltage. It can be assumed that with an increase in the voltage of the input pulse above 160 kV, the amplitude of oscillations will increase even more. However, further increasing the voltage is associated with significant technical difficulties due to the increased risk of electrical breakdown. At the same time, the calculation results have significant differences in the amplitude of oscillations:
 - the calculated oscillations are of a lower amplitude;
 - the oscillations fall-off more slowly, accordingly, the spectrum is narrower.

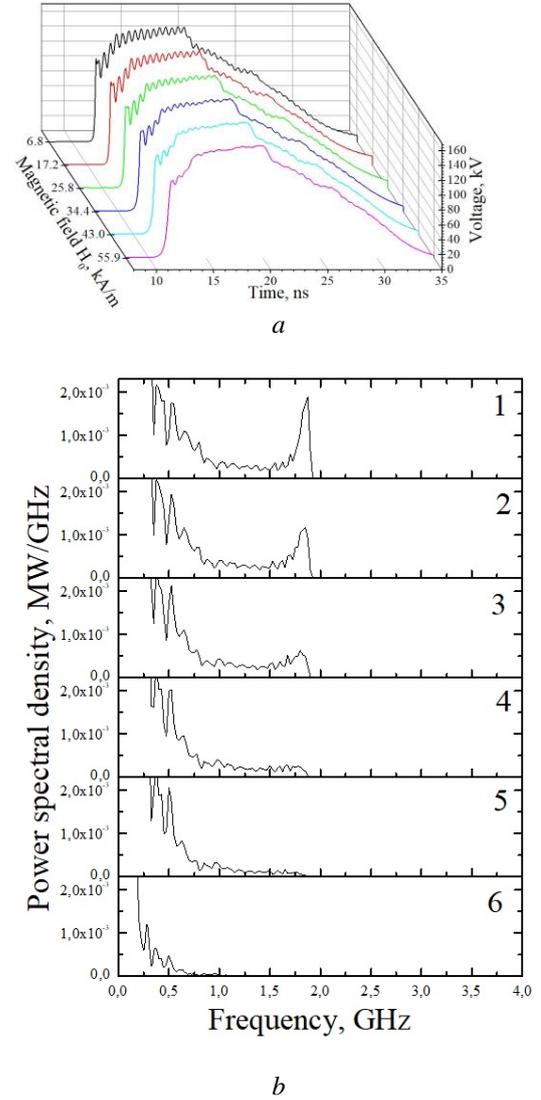

*Fig.5. The shape of voltage pulse at the output of the NLPL (sensor S3) (a) and its spectrum (b) for different magnitudes of the magnetic field $H_0$, according to the numerical calculation: 1 – 8.6 kA/m, 2 – 17.2 kA/m, 3 – 25.8 kA/m, 4 – 34.4 kA/m, 5 – 43 kA/m, 6 – 55.9 kA/m.*

Note that the frequency of the oscillations is practically independent of the magnetic field $H_0$. The dispersion curve of the quasi-TEM wave in the PS is linear at low frequencies and asymptotically approaches the straight line $f = 2$ GHz at higher frequencies (curve 1 in Fig. 6). The point of intersection of the dispersion curve of the PS and the dispersion dependence of the shock wave in the form of a straight line (2 in Fig. 6) determines the frequency of formation of oscillations of 1.9 GHz. This is the point of synchronism of the shock wave and the wave in the PS, where the phase velocities of both waves are the same, and the group velocity of the wave is less than these two velocities - i.e., the conditions for the coincidence of velocities for the Cherenkov effect are met. The dispersion curve of the wave in the PS very slowly approaches the horizontal straight line $f = 2$ GHz, so the point of intersection of the dispersion dependences depends little on the slope of the straight line. Thus, the frequency of the excited wave is determined almost entirely by the frequency properties of the PS and practically does not depend on the properties of the ferrite.

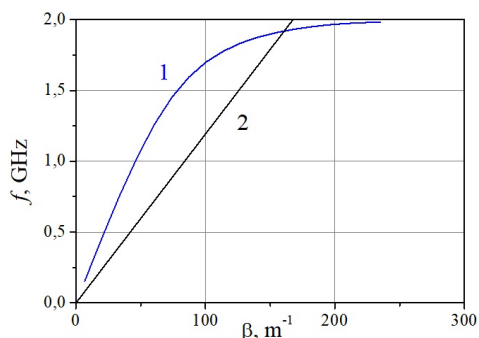

*Fig. 6. Dispersion dependences of the quasi-TEM mode of a line with PS without ferrite (1), calculated by the FDTD method, and the shock wave (pulse front) (2) obtained experimentally*

The abrupt disappearance of oscillations in the experiment at the minimum saturating magnetic field $H_0$=8.6 kA/m is also noteworthy, while in the numerical calculation they are present and do not differ from oscillations at a higher $H_0$=17.2 kA/m. According to literature data, the 200 VNP ferrites used in the experiment are characterized by a saturation magnetic field of about 14 kA/m. Thus, at the minimum magnetic field $H_0$=8.6 kA/m, the ferrite might be out of the saturation, thus altering the conditions for appearance of the oscillations associated with specific wave modes. The numerical model currently available fails taking such effects into account as it proceeds from the Landau-Lifshitz equation, which is valid near the state of full saturation for the ferromagnets.

## 4. CONCLUSION

A coaxial transmission line involving a saturated ferrite core and a periodic structure made of metal, that has been fed with a unilateral, high voltage pulse of short duration demonstrated weakly decaying VHF oscillations. The line of a 50.5 mm outer diameter, filled with an isotropic dielectric (relative dielectric constant ε = 2.25) and a set of ferrite rings with ε = 16 and μ about 4 or 5 in the saturated state, supported the oscillations at central frequencies between 1.89 GHz and 2.2 GHz, with spectral band widths about 16 percent. With magnitudes of the initial feeding pulse close to 160 kV, estimates for the instantaneous radiated power reached 20 to 22 MW.

## ЗБУДЖЕННЯ КВАЗІМОНОХРОМАТИЧНИХ ХВИЛЬ ІМПУЛЬСОМ ВИСОКОЇ НАПРУГИ У ФЕРИТОВІЙ КОАКСІАЛЬНІЙ ЛІНІЇ З ПЕРІОДИЧНОЮ СТРУКТУРОЮ

*О.Б. Батраков, С.Ю. Карелін, О.М. Лебеденко, В.С. Мухін, І.Н. Оніщенко, О.Л. Рак, В.Г. Сініцин, М.В. Воловенко*

Наведено експериментальні дані та результати чисельного моделювання щодо збудження вузькосмугових цугів хвиль гігагерцевого діапазону в коаксіальних передавальних структурах, які частково заповнені феромагнітним матеріалом і можуть включати періодично розташовані металеві вставки. Проведені експерименти підтверджують можливість збудження слабкозатухаючих електромагнітних хвиль шляхом подачі в лінію однополярних електромагнітних імпульсів високої напруги короткої тривалості. Коаксіальна лінія мала зовнішній діаметр 96 мм, заповнена ізотропним діелектриком (відносна діелектрична проникність ε = 2,25) і набором феритових кілець з ε = 16 і μ насиченого стану приблизно від 4 до 5. З піковою напругою первинного імпульсу, близької до 160 кВ, і намагнічувальним полем 17.5 кА/м, параметри збуджені хвилі досягали наступних величин: частота 1.89 ГГц; ширина смуги частот 16%; Потужність УКХ на виході близько 20 МВт.